\pdfoutput=1

\documentclass[11pt]{article}

\usepackage[preprint]{acl}

\usepackage{times}
\usepackage{latexsym}

\usepackage[T1]{fontenc}

\usepackage[utf8]{inputenc}

\usepackage[textsize=footnotesize]{todonotes} 

\usepackage{microtype}
\usepackage{graphicx}
\usepackage{subcaption}
\usepackage{booktabs}
\usepackage{amsmath}
\usepackage{multicol}
\usepackage{inconsolata}

\newcommand\boldgreen[1]{\textcolor{olive}{#1}}
\newcommand\boldred[1]{\textcolor{red}{#1}}
\renewcommand\boldgreen[1]{{#1}}
\renewcommand\boldred[1]{{#1}}

\newcommand{\shrink}{\vspace{-0.5\baselineskip}}
\newcommand{\sshrink}{\vspace{-0.25\baselineskip}}
\newcommand{\ssshrink}{\vspace{-0.125\baselineskip}}

\newcommand{\mypar}[1]{\smallskip\pagebreak[3]\noindent\textbf{#1}~}

\title{R-GPL: Hard Negative Remining for Pseudo Labelled Domain Adaptation}

\title{R-GPL: Hard Negative Remining for Generative Pseudo Labeled Domain Adaptation}

\title{R-GPL: Hard Negative Remining for Generative Pseudo Labeling for Unsupervised Domain Adaptation of Dense Retrieval}

\title{R-GPL: Remining Hard Negatives for Generative Pseudo Labeling for Unsupervised Domain Adaptation of Dense Retrieval}

\title{R-GPL: Remining Hard Negatives for Unsupervised Domain Adaptation of Dense Retrieval}

\title{R-GPL: Remining Hard Negatives for Generative Pseudo Labeled Domain Adaptation}

\title{\mbox{Remining Hard Negatives for Generative Pseudo Labeled Domain Adaptation}}

\author{Goksenin Yuksel \\
  University of Amsterdam \\ Amsterdam, The Netherlands \\ \texttt{goksenin.yuksel@student.uva.nl} \\\And
  David Rau \\
  University of Amsterdam \\ Amsterdam, The Netherlands \\
  \texttt{d.m.rau@uva.nl} \\\And
  Jaap Kamps \\
  University of Amsterdam \\ Amsterdam, The Netherlands \\
  \texttt{kamps@uva.nl} \\
  }

\begin{document}
\maketitle
\begin{abstract}
Dense retrievers have demonstrated significant potential for neural information retrieval; however, they exhibit a lack of robustness to domain shifts, thereby limiting their efficacy in zero-shot settings across diverse domains.  A state-of-the-art domain adaptation technique is Generative Pseudo Labeling (GPL). GPL uses synthetic query generation and initially mined hard negatives to distill knowledge from cross-encoder to dense retrievers in the target domain. In this paper, we analyze the documents retrieved by the domain-adapted model and discover that these are more relevant to the target queries than those of the non-domain-adapted model. We then propose refreshing the hard-negative index during the knowledge distillation phase to mine better hard negatives. Our remining R-GPL approach boosts ranking performance in 13/14 BEIR datasets and 9/12 LoTTe datasets. 
Our contributions are (i) analyzing hard negatives returned by domain-adapted and non-domain-adapted models and (ii) applying the GPL training with and without hard-negative re-mining in LoTTE and BEIR datasets.  
\end{abstract}

\shrink
\section{Introduction}
\sshrink

Dense passage retrieval \cite{dense-survey} is a key approach for text retrieval on large corpora, in the crucial first stage of modern NLP pipelines. The underlying idea is to represent queries and passages as low-dimensional vectors (also known as embeddings) to measure relevance by comparing the similarity between the query and document embeddings \cite{thakur-etal-2021-augmented}. 
Embedding queries and passages independently make these models efficient and applicable as first-stage retrievers on large corpora. Document embeddings can be pre-computed and stored, reducing computational overhead at inference time.
However, they suffer from overfitting to domain-specific training data \cite{QGen} and underperforming in novel test domains \cite{beir}. More robust approaches,  such as cross-encoder and late interaction models,  exist in the literature but come with computational and memory overheads. Therefore, they are limited in 
inference time efficiency, which is the prime concern for any practical application in production systems.

\begin{figure}[!t]
   \centering
   \includegraphics[width=0.7\linewidth]{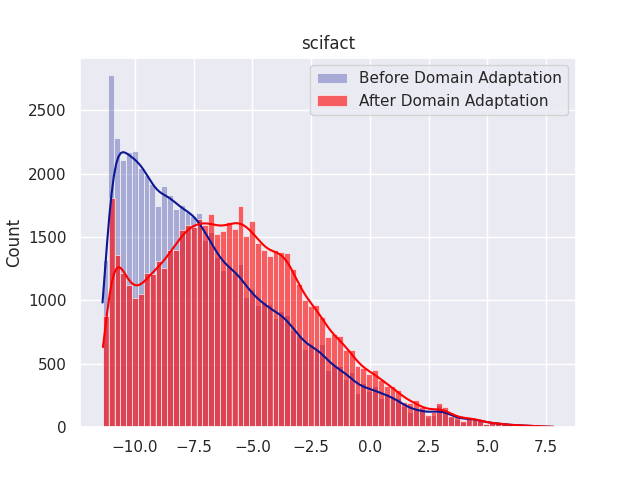}
   \caption{Score distribution of top~100 documents retrieved before and after domain adaptation. Estimated relevance scores based on the teacher model from GPL.}
   \label{teaser}
\end{figure}

\newcommand{\mainq}{Can the domain-adapted model provide more informative hard negatives to further adapt the model to the target domain?}

The lower effectiveness of dense retrieval models out-of-domain hinders the application of dense retrieval systems in zero-shot settings, making them less viable choices compared to cross-encoders or late-interaction models. To overcome this limitation, ``domain adaptation'' methodologies have been proposed.
Figure~\ref{teaser} shows the shift in score distribution between the base zero-shot model and the domain-adapted model.
Documents retrieved by the domain-adapted model are clearly more relevant to the query than those retrieved by the non-domain-adapted model. This demonstrates the potential of domain adaption of models, but also prompts the question if we can exploit this to further improve the domain adaptation.  
%
This directly motivates the main research question of this paper: \textsl{\mainq}

Unsupervised domain adaptation exploits two methods that improve performance for adapting models out-of-domain: knowledge distillation~(KD) and query generation~(QG) \cite{wang-etal-2022-gpl, QGen, beir}.
Considering that human-generated relevance judgment is quite costly  \cite{xin-etal-2022-zero, thakur-etal-2021-augmented, QGen, inpars}, robust models can be used effectively for pseudo labeling instead, and KD becomes an essential approach to improving the capacity of dense retrievers \cite{thakur-etal-2021-augmented, KD}. Given a source domain with sufficient training signals, the goal is to transfer the dense retrieval~(DR) model to a target domain without access to human editorial relevance labels. 
Query generation~(QG) models introduce synthetic training data by generating new queries to form positive pairs with documents \citep[e.g.,][]{DBLP:conf/clef/BalogAKR06}. 

Combining both procedures introduces domain-specific synthetic data to further fine-tune models in a specific domain without the need for human labels. 
GPL \citep{wang-etal-2022-gpl} has demonstrated to be a very effective non-hybrid dense retriever domain adaptation method \cite{ren2023a}. GPL distills knowledge from the cross-encoder model to the dense retrieval model using synthetic training data. To this end, GPL performs hard-negative mining step at the beginning of training, using pre-trained MSMARCO dense retrievers. The hard negatives are mined before the training and remain unchanged throughout the entire domain-adaptation process. GPL shows significant performance improvements on 18 of 19 BEIR datasets over previous methods. 

In this paper, we extend GPL by proposing to remine hard negatives during the domain adaptation process. The main idea is that instead of using the ranker that has been trained on the source domain to mine hard negatives, we employ the model undergoing the domain adaptation. During training, the model learns to retrieve more relevant hard negatives, optimizing the margin of the top-ranked documents. 
Our experiments show that remining hard negatives by the domain-adapted model during training can improve performance over using initially retrieved, static hard negatives. Our  analysis shows that the hard negatives retrieved by the domain-adapted model are significantly more relevant to the queries than pre-retrieved hard negatives. 

Our main contributions are the following.
First, we complement the original work of GPL by evaluating it on the LoTTE benchmark.
Second, we improve GPL by proposing to continuously remine hard negatives with the model undergoing domain adaptation and demonstrate performance improvements on BEIR and LoTTe benchmarks. Our method boosts the ranking performance of the domain-adapted model in 13 out of 14 datasets on BEIR and in 9 out of 12 on LoTTE.
Third, in an extensive analysis we investigate hard negatives returned by the domain-adapted retriever and show why remining hard negatives is effective for domain adaptation.

The rest of this paper is structured as follows. 
\S\ref{sec:rel} embeds our work in the literature. 
\S\ref{sec:met} details our novel domain adaptation approach based on remining hard negatives.
\S\ref{sec:exp} details our experimental setup.
\S\ref{sec:res} shows the experimental results demonstrating the retrieval effectiveness of domain adaptation.
\S\ref{sec:ana} conducts an extensive analysis how remining hard negatives impacts domain adaptation. 
The paper ends in \S\ref{sec:dis} with discussion and conclusions.
 
\shrink
\section{Related Work}
\label{sec:rel}
\sshrink

This section details related work on robust ranking models, on unsupervised domain adaptation, and on hard negative mining. 

\sshrink
\subsection{Robust Ranking Models}
\ssshrink

Cross-encoders \cite{cross-encoder} demonstrate better performance and are less prone to the out-of-domain problem \cite{beir}. They concatenate the query and passage and predict a relevance score using cross-attention. Despite their superior performance and resilience to domain shift, the requirement to concatenate the query and passage renders them less suitable for a production setting due to extremely high computational overhead.

Similarly, late interaction models have shown robustness to domain shift while delivering performance comparable to cross-encoders \cite{beir, colbert, santhanam-etal-2022-colbertv2}. These models delay the interaction between query and document tokens until the final stage. Consequently, they provide a significant speedup compared to cross-encoder models that use cross-attention from the beginning \cite{colbert}. However, late interaction models are still inefficient compared to dense retrieval models due to the need to compare every query and document token embedded in the final stage. Additionally, they consume significant memory if indexed by approximate nearest neighbor search methods like FAISS \cite{FAISS} to expedite inference \cite{macdonald-2021-single}.

\sshrink
\subsection{Unsupervised Domain Adaptation}
\ssshrink

\mypar{Query Generation} 
(QG) methods construct synthetic training data by using documents from the target domain to generate corresponding (pseudo) queries, which aims to augment the training data by generating queries that fit the target domain. QGen \cite{QGen} trains an Auto-encoder in the source domain to generate synthetic questions from a target domain document. They use binary-level relevancy labels to train the networks on generated query-document pairs. On the other hand, GPL \cite{wang-etal-2022-gpl} generate synthetic queries with a pre-trained T5 model and uses cross-encoders to label the relevancy of generated query-document pairs. This method extends the QGen method by replacing binary relevance labels with continuous relevance scores. GPL boosts the performance of the dense retriever and demonstrates strong performance in unsupervised domain adaptation for dense retrievers \cite{ren2023a}.

Similar ideas to GPL have been explored for cross-encoders. InPars ~\cite{inpars} uses a large language model (LLM) to generate synthetic queries with few-shot and zero-shot settings. They use OpenAI's public API to prompt GPT-3 to generate syntetic queries for a document, showcasing that fine-tuning cross-encoders on the synthetic data yields very strong results on different datasets. Their research demonstrates that synthetic queries provide a solid learning signal to fine-tune neural IR models. Note that \citet{inpars} did not further fine-tune dense retriever models in their experiments.

\mypar{Knowledge Distillation} 
(KD) refers to the process of transferring knowledge from a more capable model (called the teacher) to a less capable model (called the student) \cite{KD}. Applied to our setup, the goal is to improve the standard dense retriever (the student) using a more robust model (the teacher) on a given dataset. KD is a commonly used strategy in DR. GPL \cite{wang-etal-2022-gpl} and AugSBERT \cite{thakur-etal-2021-augmented} use cross-encoder to annotate unlabeled synthetic query-doc pairs to train the bi-encoder. Unlike the above methods, SPAR \cite{chen2022salient} proposes distilling knowledge from the BM25 to the DR model to transfer relevance signals from a sparse model into dense retrieval.

\sshrink
\subsection{Hard Negatives}
\ssshrink
Hard negatives are irrelevant passages with a high semantic similarity to the query. When included in the training, they have been shown to improve the ranking performance of dense retrievers by boosting the capacity to distinguish between hard cases where a passage could be relevant but is not. 

An early hard negative mining strategy is to sample lexically similar texts returned by BM25, \citet{karpukhin-etal-2020-dense} uses BM25 top documents as hard negatives. This strategy is efficient and only needs to be computed once. Hard negatives are kept fixed throughout the process. A drawback of this approach is that hard negatives are biased towards exact term matching. Later research explored dynamic hard negative mining techniques with dense retrievers. ANCE~\cite{ANCE} proposes to sample from the top retrieved texts by the optimized retriever itself, along with an asynchronous index refresh mechanism during training.  REALM~\cite{REALM} uses the dense retriever document encoder and refreshes the index of document embeddings similar to ANCE using the document embedder every k steps.  

\ssshrink
\section{Method}
\label{sec:met}
\sshrink

This section details our domain adaptation approach based on repeatedly mining hard negatives during training and domain adaptation.

\sshrink
\subsection{Dense Passage Retrieval}
\ssshrink
We follow the dense passage retrieval framework for generating embeddings for documents and queries, namely E(d) and E(q).

Given an input passage\footnote{We use passage here to refer to both query and documents.} x = {[CLS], $w_{1}$, \dots , $w_{l}$ , [SEP]}, we use the embedding model E(d), and E(q) to encode the input sequence into hidden states h =\{[CLS], $h_{1}$, \dots , $h_{l}$, [SEP]\}, where $w_{i}$ is the i-th
token; [CLS] and [SEP] are special tokens that
mark the start and end of a sentence, respectively, and $h_{i}$ is the i-th hidden state corresponding to the i-th word.
We use mean pooling over hidden states to obtain a dense representation of the entire passage. 
We ignore [CLS] and [SEP] tokens while pooling. The maximum sequence length for every model is 350.


\subsection{MarginMSE Loss}

In the following we describe the MarginMSE loss briefly. For each generated query (Q), we label the document used to generate the query as the ``relevant'' document ($D^{+}$). And the non-relevant document ($D^{-}$) is a mined hard negative. We use Margin Mean Squared Error (Margin-MSE) loss for training the models with the data triplet (Q, $D^{+}$, $D^{-}$) \cite{marginmse}. 
\shrink

{\small%
\begin{align*}
    \mathcal{L}(Q, D^{+}, D^{-}) = \text{MSE} \big( & DR(Q,D^{+}) - DR(Q,D^{-}), \\
    & CE(Q,D^{+}) - CE(Q,D^{-}) \big)
\end{align*}}%
\shrink

We calculate the margin from both the student model (DR) and the teacher model (CE). Later we use the formula below as our loss function. 

\sshrink
\subsection{GPL}
\ssshrink

GPL \citep{wang-etal-2022-gpl} is based on Generative Pseudo Labeling, a strategy in which a query generator is combined with a pseudo-label from a cross-encoder model.
GPL utilizes hard negatives to distill knowledge from the cross-encoder to the dense retriever using MarginMSE loss.

The initially selected hard negative documents are in the top-50 rankings of pre-trained retrievers. The teacher model uses these hard negatives and calculates the relevance margin between the relevant and the hard negative document to the generated query. Later, the calculated margin is utilized to force the student's margin to mimic the teacher's. This way, the student learns to embed documents and queries in latent space with similar distances to the teacher.
This forces the student (our dense ranker) to mimic the margin of the cross-encoder scores in training \cite{marginmse}. 

\citet{wang-etal-2022-gpl} showed that using pre-trained dense retrievers from MSMARCO leads to better performance improvements than BM25 hard negatives. This finding emphasizes the importance of hard negatives in the GPL framework and their effect on the training schema.

\sshrink
\subsection{R-GPL}
\ssshrink

The main idea of our approach is based on the observation that, while the base model undergoes domain adaptation, it incrementally improves at the task and language of the target domain. In our approach, we exploit this property and use the domain-adapted model to mine "better" hard negatives incrementally. To achieve this, we refresh the hard negatives with the model currently undergoing domain adaptation training every k steps.  We call our new approach R-GPL: Remining hard negatives for Generative Pseudo Labeled domain adaptation.  

In order to make our approach directly comparable to GPL, we closely follow their experimental setup and processing pipeline as described in \citet{wang-etal-2022-gpl}. 
This allows us to analyze the impact of our proposed hard negative remining approach directly to GPL's performance to understand the effect.   


\shrink
\section{Experimental Setup}
\label{sec:exp}
\sshrink

This section details our experimental setup. 

\subsection{Datasets}

BEIR \cite{beir} combines several existing datasets into a heterogeneous suite for ``zero-shot IR'' tasks, spanning bio-medical, financial, and scientific domains. We utilize open-access datasets from BEIR in our paper, which is 14 out of 19, and compare the results with recent literature.

While the BEIR datasets provide a useful test set, many capture broad semantic relatedness tasks—like citations, counterarguments, or duplicate questions–instead of natural search tasks, or else they focus on high-popularity entities like those in Wikipedia. Therefore, we extend the usual BEIR dataset and test our models on the LoTTE dataset.

LoTTE \cite{santhanam-etal-2022-colbertv2} is a new dataset for Long-Tail Topic-stratified Evaluation for IR. To complement the out-of-domain tests of BEIR, LoTTE focuses on natural user queries that pertain to long-tail topics, ones that an entity-centric knowledge base like Wikipedia might not cover. LoTTE consists of 12 test sets, each with 500–2,000 queries and 100k–2M passages. The test sets are explicitly divided by topic. The test sets include a ``pooled'' setting. The passages and queries are aggregated across all test topics to evaluate out-of-domain retrieval across a more extensive and diverse corpus in the pooled setting.

\subsection{Models}

In order to fairly compare to GPL \citep{wang-etal-2022-gpl}, we use a very similar setup with the same models. 

\mypar{BM25} For every dataset, we generate BM25 Scores using ElasticSearch engine with b = 0.75, k1 = 1.2. 

\mypar{Base} We use the zero-shot performance of the unadapted model, specifically GPL/msmarco-distilbert-margin-mse.\footnote{\url{https://huggingface.co/GPL/msmarco-distilbert-margin-mse}}

\mypar{GPL} We use the GPL model as described above.  We train GPL models on the LoTTE dataset following the exact pipeline of \citet{wang-etal-2022-gpl}. We use msmarco-distilbert-base-v3 (\textsl{Hard Negative Miner 1})\footnote{\url{https://huggingface.co/sentence-transformers/msmarco-distilbert-base-v3}} and msmarco-MiniLM-L-6-v3 (\textsl{Hard Negative Miner 2})\footnote{\url{https://huggingface.co/sentence-transformers/msmarco-MiniLM-L-6-v3}} retrievers. 

\mypar{R-GPL} We include our new remining GPL model R-GPL described above.  We run our new R-GPL model on the BEIR and LoTTE datasets, with hard-negative remining every k = 30,000 steps. For initial hard negatives, we use Hard Negative Miner 1, and Hard Negative Miner 2

\shrink
\section{Main Results}
\label{sec:res}
\sshrink

This section details our main results in terms of the retrieval effectiveness of the new R-GPL approach to domain adaptation.
\medskip 

\newcommand{\sig}{\ensuremath{^{\star}}}

\begin{table}[!t]
    \small
    \setlength\tabcolsep{0pt} 
    \begin{tabular*}{\columnwidth}{@{\extracolsep{\fill}} lcccc}
    \toprule
    \bf Corpus & \bf BM25 & \bf Base & \bf GPL  & \bf R-GPL\\
    \midrule
    \multicolumn{5}{l}{ \it Bio-Medical Information Retrieval} \\
    TREC-COVID\sig      & 68.8 & 65.1 & 71.6 & $\textbf{76.0}^{\boldgreen{\fontsize{4}{6}+4.4}}$ \\
    NF-Corpus      & \textbf{34.3} & 27.6 & 34.1 & ${34.2}^{\boldgreen{\fontsize{4}{6}+0.1}}$ \\
    \midrule
    \multicolumn{5}{l}{ \it Open Domain Question Answering (QA) } \\ 
    HotPot-QA\sig       & \textbf{60.2} & 55.4 & 56.5 & ${56.7}^{\boldgreen{\fontsize{4}{6}+0.2}}$ \\
    FIQA\sig           & 25.4 & 26.7 & 32.8 & $\textbf{33.6}^{\boldgreen{\fontsize{4}{6}+0.8}}$   \\
    NQ\sig              & 32.6 & 45.6 & 46.7 & $\textbf{50.4}^{\boldgreen{\fontsize{4}{6}+3.7}}$ \\
    \midrule
    \multicolumn{5}{l}{\it Argument Retrieval} \\ 
    Arguana        & 47.2 & 33.9 & \textbf{48.3} & ${46.4}^{\boldred{\fontsize{4}{6}-1.9}}$ \\
    WebisTouche\sig     & \textbf{34.7} & 19.6 & 23.3& ${26.4}^{\boldgreen{\fontsize{4}{6}+3.1}}$ \\
    \midrule
    \multicolumn{5}{l}{\it Duplicate Question Retrieval} \\    
    Quora\sig         & 80.8 & 81.2 & 83.2 &  $\textbf{{84.1}}^{\boldgreen{\fontsize{4}{6}+0.9}}$ \\
    CQADupstack   & 32.5 & 29.6 & 34.5 & $\textbf{{34.8}}^{\boldgreen{\fontsize{4}{6}+0.3}}$  \\
    \midrule
    \multicolumn{5}{l}{ \it Entity Retrieval} \\   
    DBPedia-Entity\sig  & 32.0 & 34.2 & 36.1 & $\textbf{{38.1}}^{\boldgreen{\fontsize{4}{6}+2.0}}$ \\
    \midrule
    \multicolumn{5}{l}{\it Citation Prediction} \\ 
    SCIDocs        & 16.5 & 13.6 & 16.1 & $\textbf{{16.2}}^{\boldgreen{\fontsize{4}{6}+0.1}}$ \\
    \midrule
    \multicolumn{5}{l}{\it Fact Checking} \\    
    Climate-FEVER  & 18.6 & 20.0 & 22.8 & $\textbf{{23.1}}^{\boldgreen{\fontsize{4}{6}+0.3}}$  \\
    FEVER\sig          & 64.9 & 76.5 & 77.9 & ${\textbf{79.1}}^{\boldgreen{\fontsize{4}{6}+1.2}}$ \\
    SCIFACT        & 69.1 & 57.1 & 66.4 & $\textbf{{67.8}}^{\boldgreen{\fontsize{4}{6}+1.4}} $ \\
    \midrule
    \textbf{Average} & 41.1 & 41.8 & 46.4 & $\textbf{{47.7}}^{\boldgreen{\fontsize{4}{6}+1.3}} $ \\
    \bottomrule
    \end{tabular*}
    \caption{Comparison of NDCG@10 results for BEIR test data. In this case MSMARCO was adapted by GPL. R-GPL updated after 30K steps. Bold Text indicates the best scoring model. The improvement points after domain adaptation are indicated. For CQADupstack, we report the mean performance of all the tasks. \mbox{}\sig  indicates statistical significance ($p < 0.05$) of R-GPL over GPL.} 
    \label{tab:beir_remine_results}
\end{table}

\begin{table}[!t]
\small
    \setlength\tabcolsep{0pt} 
    \begin{tabular*}{\columnwidth}{@{\extracolsep{\fill}} lcccc}
    \toprule
    \bf Corpus & \bf BM25 & \bf Base & \bf GPL  & \bf R-GPL \\
        \midrule
        \multicolumn{5}{l}{\it LoTTE Search Test Queries (Success@5)} \\
        Writing\sig & 63.2 & 70.6 & 77.1  & $\textbf{77.4}^{\boldgreen{\fontsize{4}{6}+0.3}}$ \\
        Recreation & 59.8 & 62.8 & $\textbf{71.0}$ & $\textbf{71.0}\phantom{^{\boldgreen{\fontsize{4}{6}\pm0.0}}}$   \\
        Science\sig & 38.6 & 46.4 & ${52.0}$ &  $\textbf{55.3}^{\boldgreen{\fontsize{4}{6}+3.3}}$ \\
        Technology & 44.5 & 57.6 &${64.1}$& $\textbf{64.9}^{\boldgreen{\fontsize{4}{6}+0.8}}$\\
        Lifestyle & 68.1 & 77.0 & ${82.8}$& $\textbf{83.4}^{\boldgreen{\fontsize{4}{6}+0.6}}$\\
        Pooled\sig& 52.4 & 62.1 & ${65.2}$ & $\textbf{67.8}^{\boldgreen{\fontsize{4}{6}+2.6}}$\\
        \midrule
        \multicolumn{5}{l}{\it LoTTE Forum Test Queries (Success@5)}\\
        Writing\sig & 66.5 & 66.8 & $72.2$ & $\textbf{73.7}^{\boldgreen{\fontsize{4}{6}+1.5}}$  \\
        Recreation\sig & 56.3 & 59.9 & $66.8$ & $\textbf{68.2}^{\boldgreen{\fontsize{4}{6}+1.4}}$ \\
        Science & 35.1 & 34.3 & $\textbf{39.7}$  & ${39.2}^{\boldred{\fontsize{4}{6}-0.5}}$ \\
        Technology\sig & 40.4 & 41.5 & $50.0$  & $\textbf{51.0}^{\boldgreen{\fontsize{4}{6}+1.0}}$ \\
        Lifestyle & 62.4 & 69.3 & \textbf{74.4} & $\textbf{74.4}\phantom{^{\boldgreen{\fontsize{4}{6}\pm0.0}}}$  \\
        Pooled\sig & 48.3 & 52.4 & $54.3$ &  $\textbf{56.7}^{\boldgreen{\fontsize{4}{6}+2.4}}$ \\
        \midrule
        \textbf{Average} & 53.0 & 58.4 & 64.5 & $\textbf{{65.2}}^{\boldgreen{\fontsize{4}{6}+0.6}} $ \\
        \bottomrule
        \end{tabular*}
\caption{Comparison of Success @5 results for LoTTE test data. In this case MSMARCO was adapted by GPL. Bold Text indicates the best scoring model. The improvement points after domain adaptation are indicated. \mbox{}\sig  indicates statistical significance ($p < 0.05$) over GPL.}
\label{tab:lotte_remine_results}
\end{table}

Table~\ref{tab:beir_remine_results} shows that R-GPL boosts the performance over the GPL model in 13 out of 14 datasets. The highest gain in improvement can be seen in the TREC-COVID dataset with 4.4 points. 
The single exception is Arguana, where both GPL and R-GPL improve the most over the Base model, with +14.4 and +12.3, respectively. 
%
We use 
a one-sided Wilcoxon signed rank test on query level NDCG@10 improvement, and find that all improvements over BM25 and Base, and most of the improvements over GPL, are significant.\footnote{We do not carry out this analysis on CQADupStack as the score is a macro average.}

Table~\ref{tab:lotte_remine_results} shows the results over the LoTTE datasets, which were not used in the earlier domain adaptation papers. 
We first observe that the GPL framework translates well to the LoTTE dataset, improving every dataset over the base model.
Similar to the BEIR results, R-GPL remining hard negatives boosts GPL's performance on model 9 out of 12 datasets, while only science forum performance degrades marginally.

\shrink
\section{Analysis}
\label{sec:ana}
\sshrink
This section provides extensive analysis of the domain adaptation, using the remining steps as a means to understand how the model adapts to the specific language of the the target domain. 

\sshrink
\subsection{Analysis Hard Negatives}
\ssshrink

\begin{figure*}[!t]
     \centering
     \begin{subfigure}[b]{.33\linewidth}
         \centering
         \includegraphics[width=\textwidth]{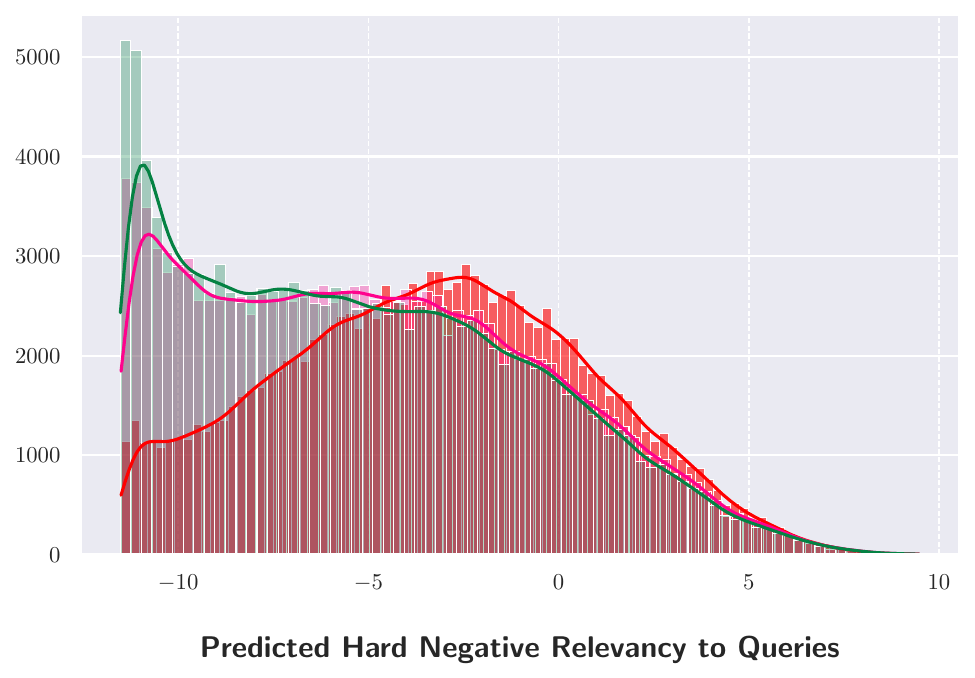}
         \caption{FIQA}
         \label{FIQA rel}
     \end{subfigure}%
    \begin{subfigure}[b]{.33\linewidth}
         \centering
         \includegraphics[width=\textwidth]{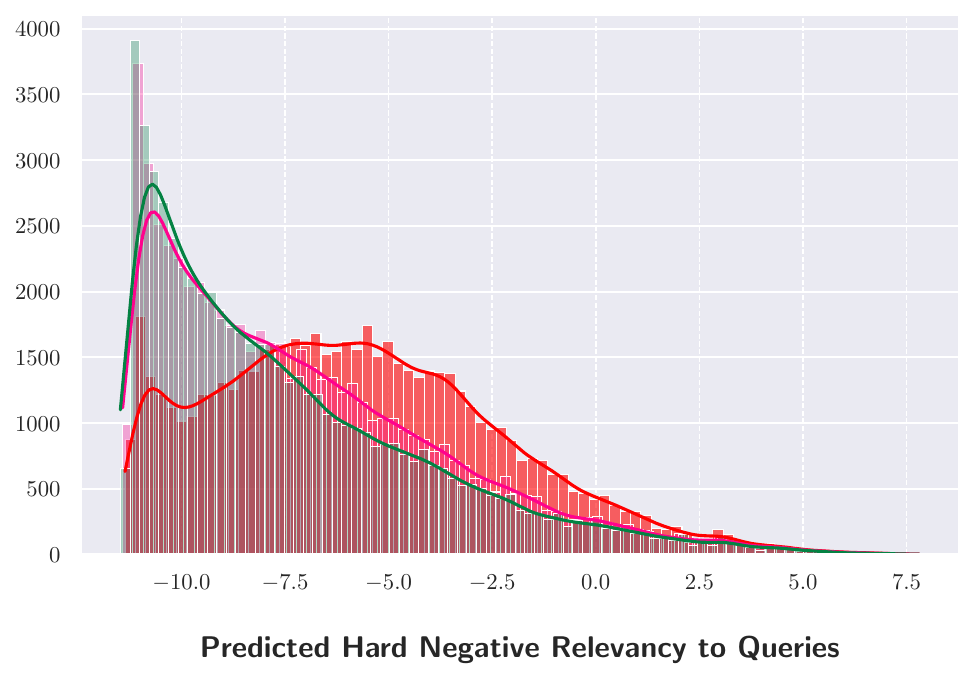}
         \caption{SCIFACT}
         \label{SCIFACT rel}
     \end{subfigure}
    \begin{subfigure}[b]{.33\linewidth}
         \centering
         \includegraphics[width=\textwidth]{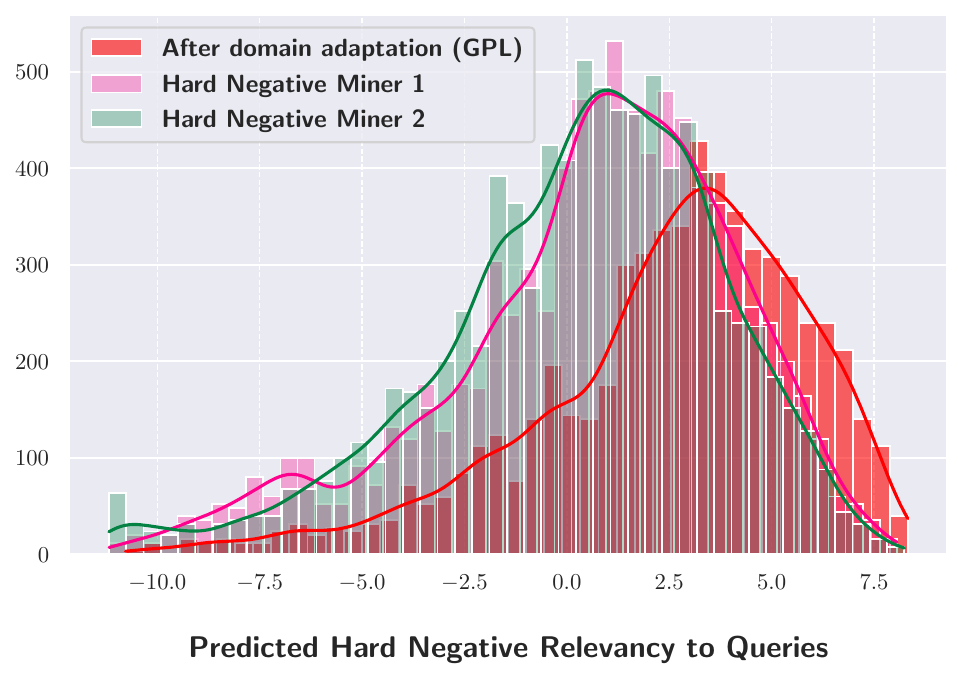}
         \caption{TREC-COVID}
         \label{SCIDOCS rel}
    \end{subfigure}
     \caption{Predicted relevancy scores for hard negatives retrieved by domain adapted model and hard negative miners. Scores are produced by the teacher model.}
     \label{RelFig}
\end{figure*}

We first analyze the hard negatives returned by the domain-adapted model and hard negative miners. We report the relevancy distribution of top-100 ranked hard negatives in Figure \ref{RelFig}.
It is encouraging to see that the domain-adapted model returns documents with higher relevancy scores in its top 100 hard negatives compared to other models.
This observation confirms that the domain-adapted model is capable of retrieving documents with higher retrieval scores. These documents are harder to distinguish from the document generating the query (hard negatives), and hence also can provide a better training signal.\footnote{Such score distributions have been studied extensively \citep[e.g.,][]{DBLP:conf/ecir/Robertson07a,DBLP:conf/ictir/ArampatzisRK09,DBLP:journals/ir/ArampatzisR11}. Theoretical and empirical analysis suggests a Normal distribution of relevant documents, and an Exponential distribution of non-relevant documents.  Hence the shift in distribution suggests that the domain-adapted model is better capable of ranking relevant documents. The exponential tail is not visible in TREC Covid, likely due to the limited ranking depth in combination with the unusually deep pooling depth used in TREC Covid.} 

\sshrink
\subsection{Remining Frequency Analysis}
\begin{figure*}[!t]
    \centering
    \begin{subfigure}[b]{.8\textwidth}
         \centering
\includegraphics[width=\textwidth]{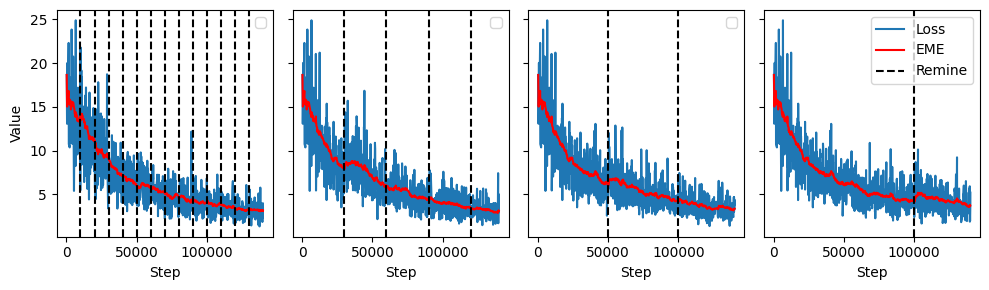}
         \caption{Distillation loss over steps for FIQA. }
         \label{fig:distill-fiqa}
     \end{subfigure}
    \begin{subfigure}[b]{.8\textwidth}
         \centering
\includegraphics[width=\textwidth]{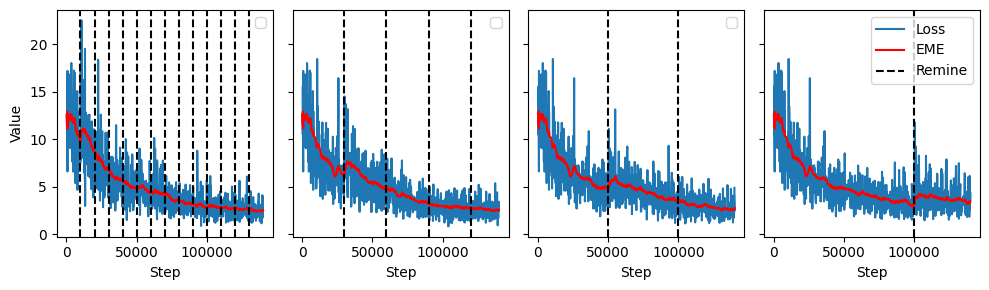}
         \caption{Distillation loss over steps for TREC-COVID.}
         \label{fig:distill-trec}
     \end{subfigure}
     \caption{Distillation training loss smoothed with an exponential moving average over length 50.
     Dashed lines indicate remined hard-negatives.}
     \label{training-loss}
\end{figure*}

\begin{figure*}[!t]
    \centering
    \begin{subfigure}[b]{\textwidth}
         \centering
\includegraphics[width=.8\textwidth]{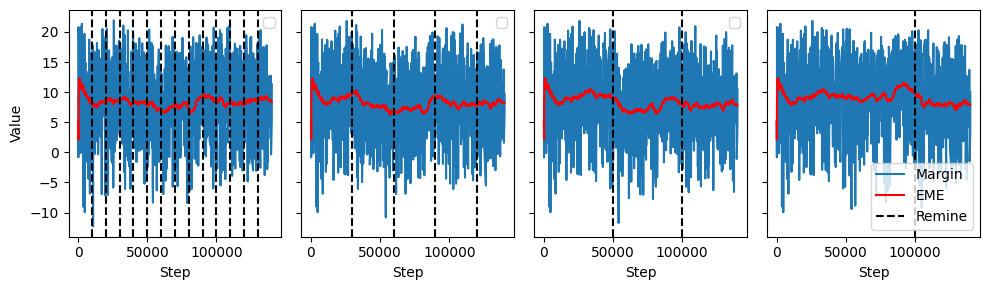}
         \caption{FIQA}
         \label{fig:margin-fiqa-train}
     \end{subfigure}
    \begin{subfigure}[b]{\textwidth}
         \centering
\includegraphics[width=.8\textwidth]{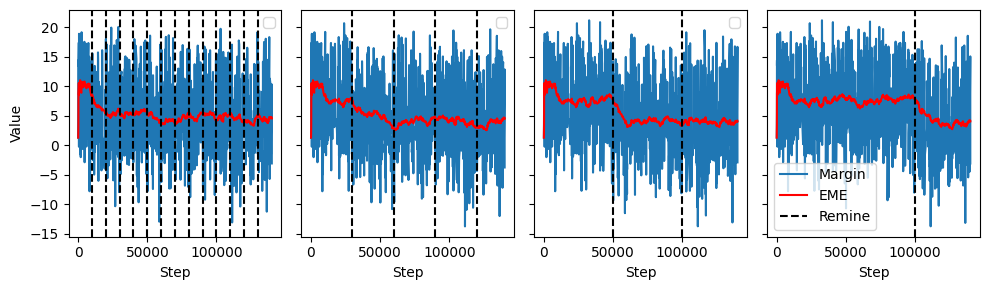}
         \caption{TREC-COVID}
         \label{fig:margin-trec-train}
     \end{subfigure}
     \caption{Predicted cross encoder (CE) relevance margin between query (Q), relevant document $(D^{+})$, and hard negative document $(D^{-})$. The margin is calculated using CE(Q,$D^{+}$) - CE(Q,$D^{-}$). }
     \label{training-margin}
\end{figure*}

\ssshrink
We alter the remining frequency value (k), and apply the R-GPL on the datasets analyzed in Figure~\ref{RelFig}.  We also report the training loss and teacher margin during the training in Figures~\ref{training-loss} and~\ref{training-margin}.

\begin{table}[!t]
    \small
    \setlength\tabcolsep{0pt} 
    \begin{tabular*}{\columnwidth}{@{\extracolsep{\fill}} lccccc}
    \toprule
    \bf Corpus & \bf GPL & \bf 10k & \bf 30k & \bf 50k & \bf 100k\\
    \midrule
    TREC-COVID   & $71.6$  & 73.7 & \textbf{76.0} & 75.1 & 75.5 \\
    FIQA     & $32.8$ & \textbf{34.0} & 33.6 & 33.8 & 33.5 \\ 
    SCIFACT    & 66.4 & 66.1 & \textbf{67.8} & 67.3  & 67.7 \\
    \bottomrule
    \end{tabular*}
    \caption{Comparison of NDCG@10 results for TREC-COVID, FIQA, and SCIFACT datasets. GPL Xk depicts the model with hard negative remining every X step.}
    \label{tab:reminingtab}
    \end{table}

Table \ref{tab:reminingtab} depicts that hard-negative re-mining with 30,000 steps yields the best performance improvements. We observe improvements over the plain GPL model in all datasets regardless of the k value.

Figure~\ref{fig:distill-fiqa} and Figure~\ref{fig:distill-trec} show that distillation loss increases after mining with hard negatives and later settles down to previous values.  However, this effect is only visible when we have remined the hard negatives for the first time. Later, we do not observe the increase in distillation loss. Furthermore, as the training continues, loss decreases, suggesting that the dense retriever model learns to mimic the margin of the cross-encoder.

Figures~\ref{fig:margin-fiqa-train}, and~\ref{fig:margin-trec-train} depict the predicted margin of the teacher. These figures show that after mining the hard negatives, the margin drops. This phenomenon is especially visible after the first refresh. 
Our analysis shows that the distillation loss of the student model (our R-GPL model) is highly coupled with the teacher's margin. 
\vspace*{-0.05\baselineskip}
Figure ~\ref{fig:margin-fiqa}, and ~\ref{fig:margin-trec} depicts the average relevance scores over hard negative mining steps. After remining the hard negatives for the first time, we see a sharp increase in relevancy. Later these relevancies increase gradually by each remining step. 

\sshrink
\subsection{Query Document Embeddings}
\ssshrink

\begin{figure}[!t]
    \centering
    \begin{subfigure}[b]{0.45\textwidth}
         \centering
\includegraphics[width=0.8\textwidth]{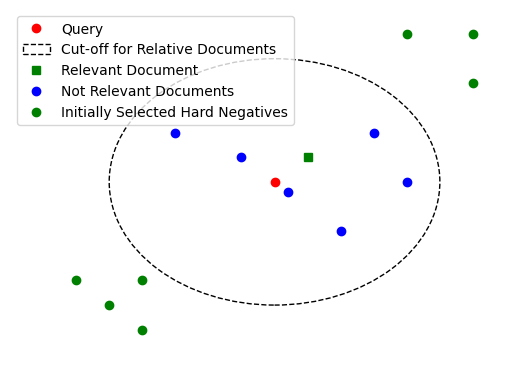}
         \caption{Initially mined hard negatives have shifted outside the vicinity of query embedding due to the domain adaptation.}
         \label{fig:before}
     \end{subfigure}
    \begin{subfigure}[b]{0.45\textwidth}
         \centering
\includegraphics[width=0.8\textwidth]{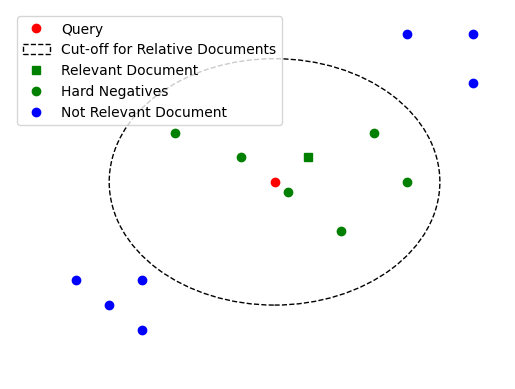}
         \caption{Embedding space right after the hard negative remining. Hard negatives are selected from the vicinity of the query.}
         \label{fig:after}
     \end{subfigure}
     \caption{2D projection of query and document embeddings used in GPL Framework.}
\end{figure}

Figure~\ref{fig:before} and Figure~\ref{fig:after} demonstrate that embeddings of static hard-negatives shift outside of the query's proximity. Refreshing the hard negatives selects the documents that are in proximity to the query embedding.

\shrink
\section{Discussion and Conclusions}
\label{sec:dis}
\sshrink

\begin{figure}[!t]
    \centering
    \begin{subfigure}[b]{0.38\textwidth}
         \centering
\includegraphics[width=\textwidth]{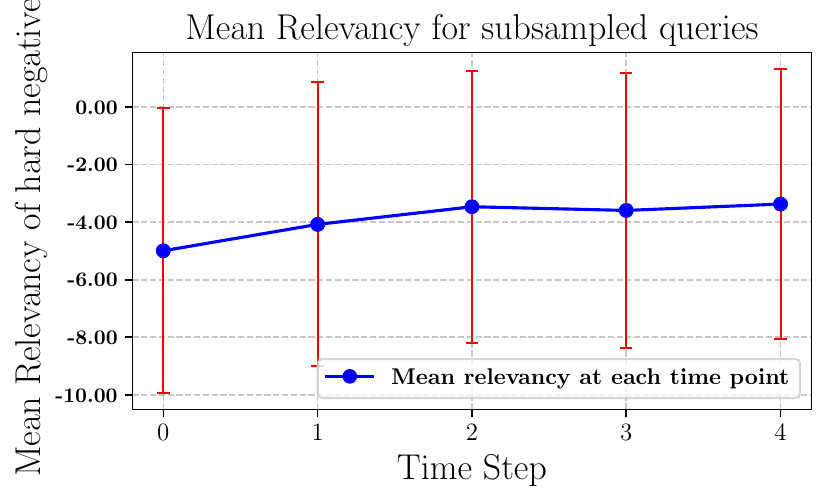}
         \caption{FIQA}
         \label{fig:margin-fiqa}
     \end{subfigure}
    \begin{subfigure}[b]{0.38\textwidth}
         \centering
\includegraphics[width=\textwidth]{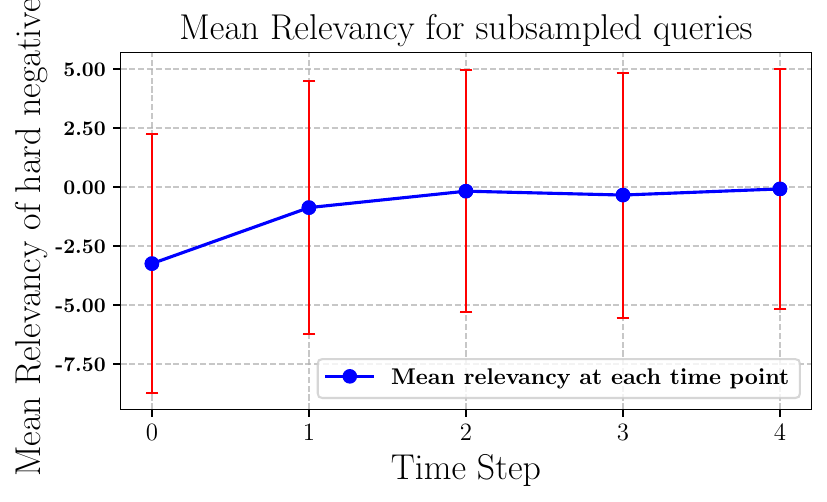}
         \caption{TREC-COVID}
         \label{fig:margin-trec}
     \end{subfigure}
     \caption{Mean relevancy of returned hard negatives for FIQA and TREC-COVID. Time step 0 indicates the initial hard negatives, each time step after 0 indicates the hard negative remining. Error bars indicate the standard deviation of relevancies at time point t. 1,000 generated queries were randomly selected, and the hard negative relevancies were calculated using the teacher cross-encoder. For each query, top 50 hard negatives are used. }
\end{figure}


The key motivation of this paper is the observation that documents retrieved by an domain-adapted model have higher relevance scores than those of the base model.
As we perform domain adaptation with such relevance scores as pseudo labels, the dense retriever can retrieve more relevant documents from the target domain. Moreover, a domain-adapted retriever can also return more relevant hard negatives, that present a better training signal to further improve domain adaptation.
We proposed using a hard negative remining strategy R-GPL, and showed that this leads to a better adaptation to the target domain.

Figures~\ref{fig:margin-fiqa} and~\ref{fig:margin-trec} indicate that by refreshing the index of hard negatives, we train the dense retriever model with more relevant hard negatives compared to the static training schema. This training schema outperforms the one without the hard negative remining in most test sets. One plausible explanation for this is the optimization of the top-ranked documents. When we have fresh hard negatives from the model, we optimize the margin of top-ranked documents. Figure~\ref{fig:before} and Figure~\ref{fig:after} demonstrate this phenomenon.  When hard negatives are refreshed, the training examples cover the proximity of the query embeddings. Moreover, it is possible that with hard negative refreshes, the model trains with a broader context. With static hard negative mining, we keep the identical document IDs in the training set; however, with the refreshes, those document IDs change, thus covering the broader document set \cite{ANCE}.  This finding is also endorsed by findings from Table~\ref{tab:reminingtab}, showing that remining hard negatives, regardless of the frequency, boosts the domain adaptation performance.

Another interesting finding is the significant increase in the relevance of the hard negatives after the first refresh. After remining the hard negatives with the domain-adapted model, we get novel and higher-scoring documents for the generated query. 
This could reveal the reason for the significant peak in loss seen in Figures \ref{fig:distill-fiqa} and \ref{fig:distill-trec}. The dense retriever fails to match the teacher's margin with the fresh hard negatives, resulting in the peak. As training progresses, loss gradually keeps decreasing, showing that the dense retriever learns to embed new documents. This finding possibly results in less prominent peaks over the training, showcasing that the dense retriever learns to mimic the margin of the teacher even with new documents.

\medskip 



Recapitulating, this paper makes a number of contributions. 
First, we introduce R-GPL by proposing to continually remine hard negatives while the model is undergoing the domain-adaptation in training. 
Our model extends the original work of GPL \cite{wang-etal-2022-gpl}, which this paper also evaluates on a second benchmark LoTTE.
Second, we observe that R-GPL performs well on both BEIR and LoTTE, outperforming zero-shot ranking and the original GPL. Compared to GPL, our method boosts the ranking performance of the domain-adapted model in 13 out of 14 datasets on BEIR and in 9 out of 12 on LoTTE.
Third, we conduct an extensive analysis of the hard negatives returned by the domain-adapted retriever, showing why remining hard negatives is effective for domain adaptation.

More generally, our findings underscore the crucial role of high-quality hard negatives in IR research and open up further possibilities for using more efficient dense retriever models in inference times on out-of-domain settings. 
Such as ranking component is typically the first stage of a modern NLP pipeline, in any realistic application setting over large corpora.  Unsupervised domain adaptation is an attractive approach to be able to any domain of application, where labeled training data at scale is typically not available.  Improving the first stage of an NLP pipeline can of immense value, as this stage acts as a gate-keeper determining what documents or passages are even considered for downstream complex NLP processing.

\section{Limitations}

Although we used a diverse set of 26 tasks and corpora, our research is restricted to English corpora and common document genres.  It is important to further investigate the effectiveness of domain adaptation approaches for cross-language or multi-language and multi-modal settings, and on more diverse tasks and corpora.  In fact, it is known that zero shot approaches as well as larger LLMs exhibit impressive performance on common tasks and topics, but can perform far less impressive on highly specific tasks and domains.  We believe that domain adaption approaches can be a fruitful approach to study such shifts over language and document genres, and yields insights to further improve out-of-domain performance. 

Our approach to remining hard negative increases the training complexity of models.  Classic lexical retrieval models such as BM25 remain unsurpassed in terms of training and inference time efficiency, due to sparse representations and highly efficient indexing approaches scaling to billions of documents. Dense retrievers, as studied in this paper, exhibit still attractive inference time complexity due to efficient indexing of document embeddings. In a production setting, with possibly millions of requests at query time, inference time complexity is far more important than training efficiency at indexing time, done only once, in terms of the carbon footprint of NLP. 

Although the same inference time complexity, domain adaption increases the training complexity compared to zero shot dense retrievers.  
%
We have proposed synchronously refreshing the index of hard negatives with respect to the training. This training schema adds a computational overhead, especially for large corpus sizes. We find that domain adaptation time increases for large corpora (> 1M Documents) compared to domain adaptation without remining hard negatives. However, at the cost of additional training for the model, we have a higher quality for the same inference time (which is the main concern in real-world applications). To further reduce the training efficiency, in future work we will investigate to perform this training schema asynchronously as done by \citet{ANCE}.  


\bibliography{acl_latex}

\appendix


\newcommand{\anon}[1]{#1}
\renewcommand{\anon}[1]{\textsl{withheld to preserve anonymity}}

\end{document}